\documentclass[traditabstract]{aa} 
\usepackage{graphicx}
\usepackage{natbib}
\usepackage{txfonts}
\begin{document}
\title{A 5.5-year robotic optical monitoring of Q0957+561: substructure in a non-local cD galaxy}

\author{V. N. Shalyapin\inst{1,2} 
\and
L. J. Goicoechea\inst{1}
\and
R. Gil-Merino\inst{1}}

\institute{GLENDAMA Team, Universidad de Cantabria, Avda. de Los Castros s/n, 39005 Santander, 
Spain\\
\email{vshal@ukr.net, goicol@unican.es, r.gilmerino@gmail.com}
\and
Institute for Radiophysics and Electronics, National Academy of Sciences of Ukraine, 12 Proskura 
St., 61085 Kharkov, Ukraine}


\abstract{New light curves of the gravitationally lensed double quasar Q0957+561 in the $gr$ 
bands during 2008--2010 include densely sampled, sharp intrinsic fluctuations with unprecedentedly
high signal-to-noise ratio. These relatively violent flux variations allow us to very accurately 
measure the $g$-band and $r$-band time delays between the two quasar images A and B. Using 
correlation functions, we obtain that the two time delays are inconsistent with each other at 
the 2$\sigma$ level, with the $r$-band delay exceeding the 417-day delay in the $g$ band by about 
3 days. We also studied the long-term evolution of the delay-corrected flux ratio $B/A$ from 
our homogeneous two-band monitoring with the Liverpool Robotic Telescope between 2005 and 
2010\thanks{Tables 1 and 2 corresponding to the Liverpool Robotic Telescope light curves are 
only available in electronic form at the CDS via anonymous ftp to cdsarc.u-strasbg.fr 
(130.79.128.5) or via http://cdsweb.u-strasbg.fr/cgi-bin/qcat?J/A+A/.}. 
This ratio $B/A$ slightly increases in periods of violent activity, which seems to be 
correlated with the flux level in these periods. The presence of the previously reported dense 
cloud within the cD lensing galaxy, along the line of sight to the A image, could account for 
the observed time delay and flux ratio anomalies.}


   \maketitle

\section{Introduction}
The optical continuum variability of the gravitationally lensed double quasar \object{Q0957+561} 
at redshift $z$ = 1.41 has been widely studied since its discovery by \citet{Wal79}. Several 
monitoring campaigns focused on the determination of the time delay between the two quasar 
images A and B \citep[e.g.,][]{Van89,Kun97,Ser99}, where a major breakthrough occurred in 
\citet{Kun97}, who used Apache Point Observatory (APO) data. The 1.5-year monitoring programme with 
the APO 3.5 m telescope led to an accurate time delay $\Delta t_{BA}$ = 417 $\pm$ 3 d (2$\sigma$ 
confidence interval; A leading) in the $g$ band. \citet{Kun97} also reported $\Delta t_{BA} 
\sim$ 420 d in the $r$ band, which was consistent with the $g$-band delay measurement. A recent 
2.5-year campaign with the Liverpool 2 m robotic telescope (LRT) has confirmed the APO $g$-band 
delay, but it has not allowed us to measure a reliable time delay in the $r$ band \citep[][Paper 
I]{Sha08}. Unfortunately, very accurate estimates of multiband delays between 
\object{Q0957+561A} and \object{Q0957+561B} remain elusive because of the absence of very 
prominent flux variations with signal-to-noise ratio $S/N \geq$ 10, where $S/N$ for a given 
fluctuation is defined as the ratio between its semiamplitude and mean flux error (see Paper I). 
The strong gravitational lensing scenario predicts the existence of an achromatic delay 
\citep[e.g.,][]{Sch92,Koc04}, while the possible detection of different delays in different 
optical bands would provide extremely valuable information on the physical properties of the 
intervening medium. 

Using the APO light curves for the two quasar images, \citet{Col01} found that the $r$-band main 
fluctuations lag with respect to those in the $g$-band by 3.4 $^{+1.5}_{-1.4}$ d (1$\sigma$ 
interval). This interband delay was interpreted as clear evidence for stratified reprocessing 
within an accretion disc that is irradiated by a central high-energy source. The accretion disc 
would orbit the central supermassive black hole of the quasar. Interestingly, $\Delta t_{rg} 
\sim$ 4 d for the image B data alone, whereas $\Delta t_{rg} \sim$ 1 d for the image A data 
alone. These two estimates agreed within the 1$\sigma$ error bars, but the shortest delay from A 
data was thought to be underestimated as a result of the relatively poor sampling and 
variability behaviour \citep{Col01}. The LRT follow-up of \object{Q0957+561A} also led to 
interband delay estimates centred on 3--4.5 d (Paper I), seemingly supporting the four-day value 
for both images. We note that the presence of equal interband delays for the two quasar images 
is equivalent to the occurrence of equal delays between images in different bands.  

One can also obtain the delay-corrected flux ratio at time $t$: $B/A = S_B(t)/S_A(t-\Delta 
t_{BA})$, where $S_A$ and $S_B$ are fluxes of \object{Q0957+561A} and \object{Q0957+561B}, 
respectively. Although the strong gravitational lensing scenario produces achromatic and 
stationary flux ratios of lensed quasars \citep[e.g.,][]{Sch92,Koc04}, actual scenarios are not 
so simple. Chromatic flux ratios are usually related to differential extinction 
\citep[e.g.,][]{Fal99,Eli06} or differential microlensing \citep[e.g.,][]{Yon08}. In addition, 
time-variable flux ratios are likely due to differential microlensing by stars in lensing 
galaxies \citep[e.g.,][]{Irw89,Par06}. The light rays associated with the two images of 
\object{Q0957+561} pass through two separate regions within the central cluster cD galaxy at $z$ 
= 0.36 acting as main gravitational lens \citep{Sto80,You80,Gar92}. Thus, while the optical 
continuum of the B image probably does not suffer significant dust extinction, the optical 
continuum light of the A image is affected by a dense dusty cloud inside the cD galaxy 
\citep{Goi05a,Goi05b}. This differential extinction produces a chromatic flux ratio, whose 
$R$-band value was basically constant from 1987 through 2000 \citep[e.g., see Fig. 3 
of][]{Osc02}. The LRT observations also support the constancy of $B/A$ over the 2000--2007 
period in the $g$ and $r$ bands. Despite the fact that stars in the main lensing galaxy may 
induce variations in $B/A$ (see above), the time-domain studies of \object{Q0957+561} over two 
decades failed to detect these variations. 

We conducted a long-term photometric monitoring programme of \object{Q0957+561} using the LRT at 
La Palma, Canary Islands. The observations are part of the Liverpool Quasar Lens Monitoring 
(LQLM) project \citep{Goi10}. Two-colour light curves of \object{Q0957+561} during the first 
phase of this project (LQLM I; from January 2005 to July 2007) were published in Paper I. Here, 
in Sect. 2, we present new light curves in the $g$ and $r$ bands (LQLM II; from February 2008 to 
July 2010). These new light curves show densely sampled, sharp intrinsic fluctuations with $S/N 
\sim$ 10, which are used in Sect. 3 to measure delays with unprecedented accuracy. In Sect. 4, 
we discuss the time-evolution of $B/A$ over the 2000--2010 decade in the $g$ and $r$ bands. Our 
main conclusions are included in Sect. 5. In Sect. 6, we briefly comment on some scenarios that
could account for the observational results, as well as future prospects. 
 
\section{Observations and data reduction}

All LQLM II optical frames of \object{Q0957+561} were obtained with RATCam. This is a CCD camera 
with a $4\farcm6\times4\farcm6$ field of view, having a pixel scale of $\sim 0\farcs27$ (binning 
2$\times$2). To obtain a photometric signal-to-noise ratio of $\sim$ 100 for the two quasar 
images for each observing night, we set the exposure times to 120 s per night in the $g$ and $r$ 
bands. Apart from the basic pre-processing tasks included in the LRT pipeline, we cleaned some 
cosmic rays and interpolated over bad pixels using the bad pixel mask. 

The pre-processed frames flow through our photometric pipelines to subsequent stages of 
processing \citep[e.g., see the flowchart in Figure 1 of][]{Goi10}. At an initial stage, the 
crowded-field photometry pipeline produces the instrumental fluxes of the quasar images. The 
frames that are of little or no interest were then removed from the initial data set. In Paper I 
we showed that quasar images with signal-to-noise ratio above 80 produce high-quality 
photometric results. Thus, only $gr$ frames with signal-to-noise ratio $\geq$ 80 over 
\object{Q0957+561A} are passed through the transformation pipeline. This pipeline transforms 
instrumental magnitudes into SDSS magnitudes, and the calibration-correction scheme is described 
in Appendix A of Paper I. 

For the long-term data in the $g$ and $r$ bands, we used a sophisticated transformation model 
incorporating zero-point, colour and inhomogeneity terms. Although this last term played an 
important role when analysing earlier observations with the LRT (2005--2007; Paper I), the new 
data in the 2008--2010 period indicate that inhomogeneities have been weaker in the most recent 
years. Some improvements to the telescope in September 2007 seem to have decreased 
inhomogeneities and typical seeing values. To obtain $g$-SDSS magnitudes of a quasar image, we 
initially considered an average colour $\langle (g - r)_{SDSS} \rangle$ in its colour correction 
(see Appendix A of Paper I). However, variations of $(g - r)_{SDSS}$ may introduce a 
non-negligible colour noise, which should be removed from the brightness record. The amplitude 
of the colour noises for the two images is $\sim$ 10 mmag, and we eliminated these systematic noises 
in our 2008--2010 $g$-SDSS records. We also turned magnitudes into fluxes (in mJy) using SDSS 
conversion equations\footnote{http://www.sdss.org/dr7/algorithms/fluxcal.html.}. 

   \begin{figure}
   \centering
   \includegraphics[angle=0,width=7cm]{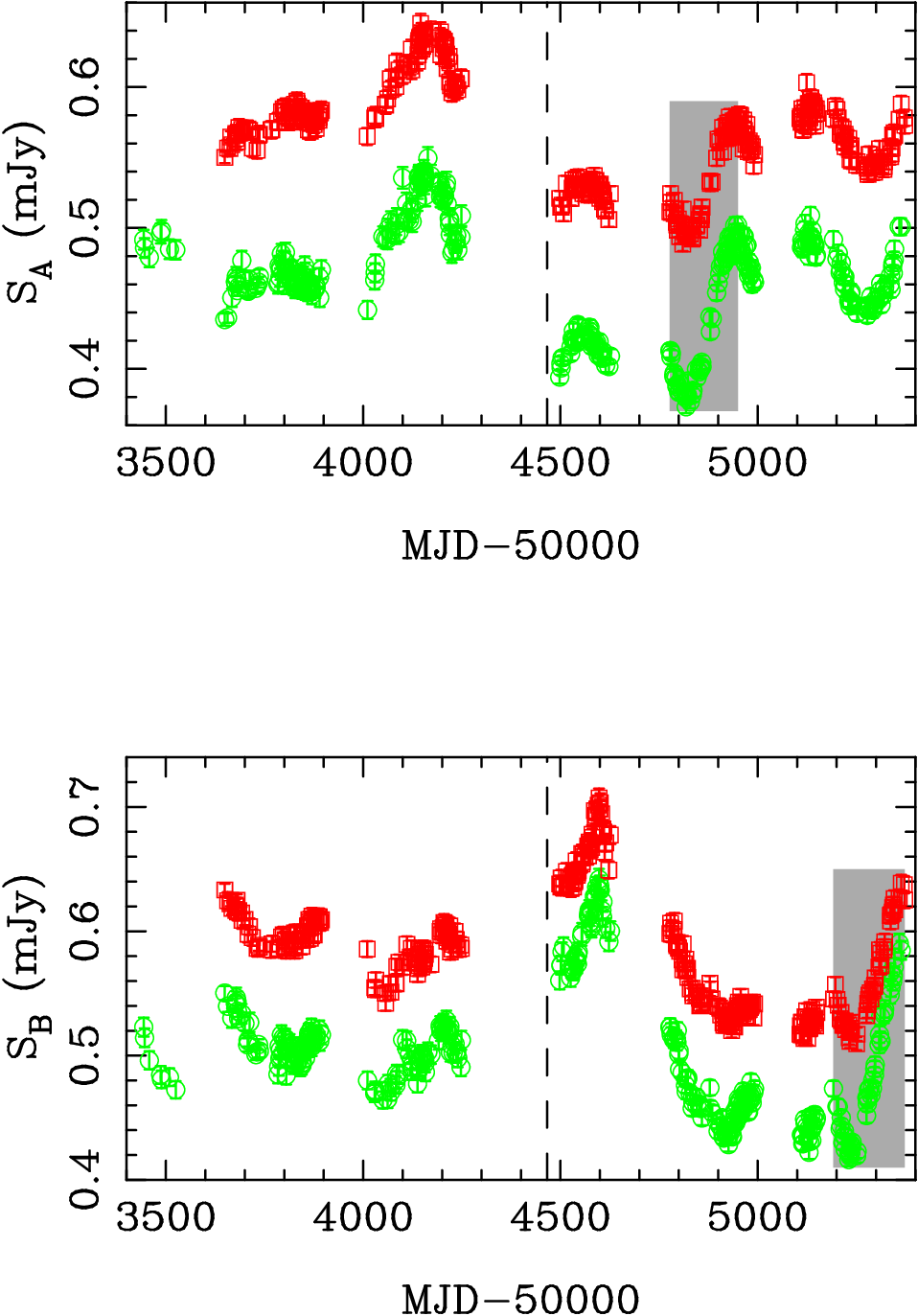}
   \caption{Fluxes of Q0957+561A (top panel) and Q0957+561B (bottom panel) in the $g$ and $r$ 
   bands of the SDSS photometric system. Circles denote the $g$-SDSS light curves and squares 
   represent the $r$-SDSS records. The whole optical data set is separated into two parts by a 
   vertical dashed line: LQLM I (2005--2007) and LQLM II (2008--2010). The new effort in the 
   2008--2010 period doubles our contribution to the two-colour variability database of 
   Q0957+561. We also highlight the LRT main fluctuations using grey rectangles (see main 
   text).}
   \label{figopt}
   \end{figure}

   \begin{figure*}
   \centering
   \includegraphics[angle=-90,width=15cm]{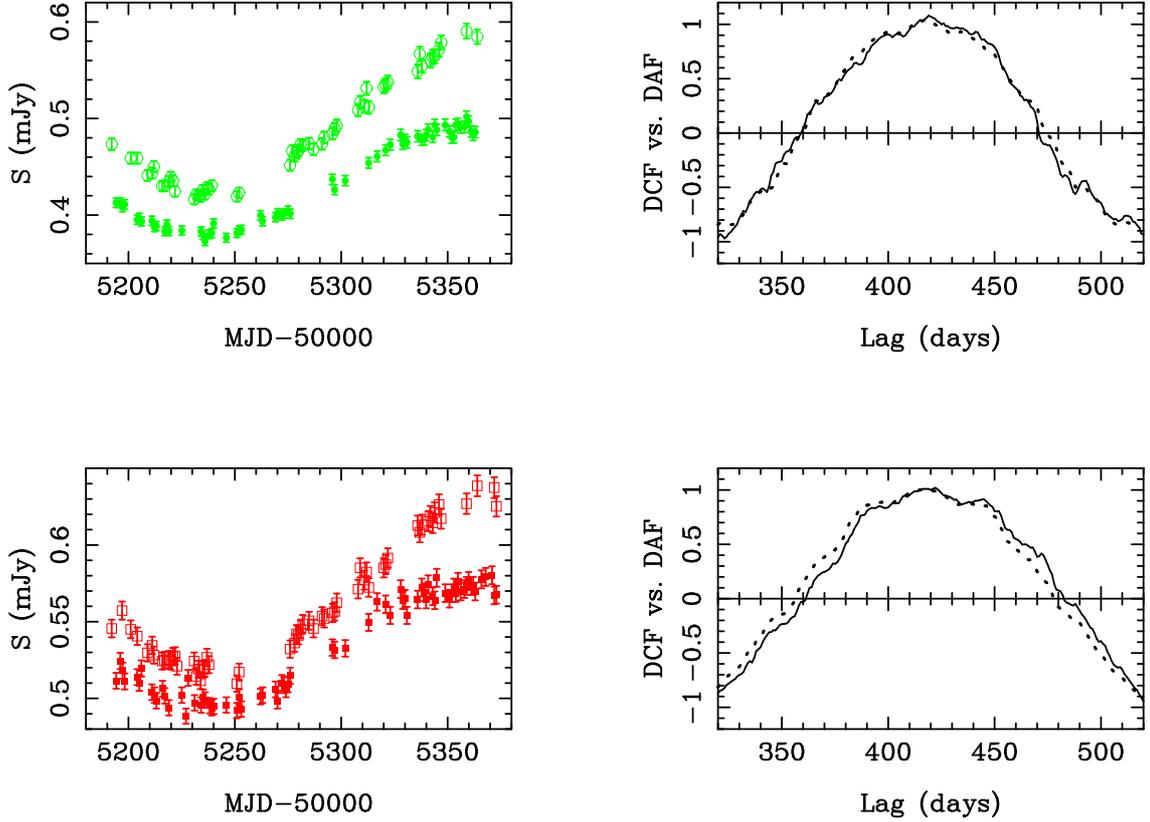}
   \caption{LRT main fluctuations and some of their correlation functions. {\it Top left 
   panel}: the fluxes of Q0957+561A (shifted forwards in time in 417 d; filled circles) and 
   Q0957+561B (open circles) in the $g$ band. {\it Top right panel}: the AB cross-correlation 
   (solid line), and the average of the AA and BB autocorrelations (shifted by + 417 d; dashed 
   line) in the $g$ band for $\alpha$ = 6 d. {\it Bottom left panel}: the fluxes of Q0957+561A 
   (shifted forwards in time in 417 d; filled squares) and Q0957+561B (open squares) in the $r$ 
   band. {\it Bottom right panel}: the AB cross-correlation (solid line), and the average of 
   the AA and BB autocorrelations (shifted by + 417 d; dashed line) in the $r$ band for $\alpha$ 
   = 9 d.}
   \label{figdel}
   \end{figure*}

A large optical variability database, incorporating previous LQLM I fluxes and the new LQLM II 
light curves of \object{Q0957+561}, is available in tabular format at the CDS\footnote{Pipeline 
outputs, magnitudes and fluxes are also publicly available in the LQLM data-tools releases at 
http://grupos.unican.es/glendama/.}: Table 1--2 include 357 $g$-SDSS and 371 $r$-SDSS pairs of 
fluxes ($S_A$,$S_B$), respectively. Each of these tables contains the following information. 
Column 1 lists the observing date (MJD--50\,000), Columns 2 and 3 indicate the flux and its 
error for the image A, and Columns 4 and 5 give the flux and its error for the image B. The LQLM 
optical light curves of \object{Q0957+561A} (top panel) and \object{Q0957+561B} (bottom panel) 
are shown in Fig.~\ref{figopt}. A vertical dashed line on 1 January 2008 separates the LQLM I 
and II periods. In the second monitoring period, there are 215 $g$-SDSS fluxes for each quasar 
image (circles), as well as 239 $r$-SDSS pairs of fluxes (squares). We achieve 1--1.3\% 
photometric accuracy during the 2008--2010 period. Moreover, excluding the unavoidable seasonal 
gaps, the average separation between adjacent data is only three days. The new LRT light curves 
display four densely sampled, very prominent variations (see the two grey highlighted regions in 
Fig.~\ref{figopt}). The two variations in $S_A$ are basically repeated in $S_B$ 14 months later, 
which means that these four fluctuations have an intrinsic origin. They also have $S/N \sim$ 10, 
and can be considered as the LRT main fluctuations.

\section{Time delays from the LRT main fluctuations}

Although a $g$-band time delay between quasar images of about 417 d is now firmly established, 
it is based on intrinsic flux variations with 3 $< S/N <$ 7 (see Paper I and references 
therein). Hence, the new $g$-band features within the grey rectangles in Fig.~\ref{figopt} 
(circles) represent a unique opportunity to measure the $g$-band delay with a very low 
uncertainty. These two well-sampled fluctuations with $S/N$ = 13 are drawn together in the 
top left panel of Fig.~\ref{figdel}, where $S_A$ (filled circles) is shifted 417 d forwards in 
time. The time delay is determined by comparing the discrete cross-correlation function ($DCF$) 
to the discrete autocorrelation function ($DAF$). More properly, the delay corresponds to the 
minimum of the square difference between the $DCF$ and the time-shifted $DAF$. This is the 
$\delta^2$ technique \citep[e.g.,][]{Ser99} relying on discrete correlation functions 
\citep{Ede88}. The $\delta^2$ minimisation is a non-parametric method, which implies that 
one does not {\it a priori} assume a chosen model to relate the shapes of $S_A$ and $S_B$ (see 
below).

The differences $S_A - \langle S_A \rangle$ and $S_B - \langle S_B \rangle$ are the key pieces 
in the $DCF$, therefore we resampled both $S_A$ and $S_B$ to obtain two curves with similar sampling 
(45 data points each), and to avoid biases between the averages $\langle S_A \rangle$ and $\langle 
S_B \rangle$. We evaluate the discrete correlation functions every day in two wide ranges of 
lags including correlation and anti-correlation peaks. The $DCF$ and $DAF$ were binned in 
2$\alpha$ intervals centred at the lags, where $\alpha \leq$ 10 d. For $\alpha$ = 3 d, both 
functions are very noisy, whereas for bin semisizes of 6 or 9 d, the discrete correlation 
functions in the $g$ band have a smoother behaviour. In the top right panel of 
Fig.~\ref{figdel}, using $\alpha$ = 6 d, we show the $DAF$ shifted by + 417 d (dashed line) and
the $DCF$ (solid line). As expected, the two trends agree very well. We 
also followed a standard Monte Carlo approach to generate 1000 synthetic data sets and determine 
time delay errors. In each synthetic light curve, the observed fluxes were modified by random 
Gaussian deviations that are consistent with the measured uncertainties. We applied the $\delta^2$ 
minimisation (see above) to each synthetic data set, and thus obtain 1000 delays for each value 
of $\alpha$. Through the distributions of delays for bin semisizes of 6 and 9 d, our final 
$g$-band measurement is $\Delta t_{BA}$ = 416.5 $\pm$ 1.0 d (1$\sigma$ interval). We also obtained
the constraint: $\Delta t_{BA} <$ 418.5 d at the 99\% confidence level.

We repeated the procedure described in the two previous paragraphs, but using the $r$-band 
data in the bottom left panel of Fig.~\ref{figdel} instead of those in the $g$ band. In this 
panel, the fluxes of the A image (filled squares) are shifted 417 d forwards in time. For 
$\alpha$ values of 6 or 9 d, the corresponding $DCF$ and $DAF$ are reasonably smooth. For 
example, the bottom right panel of Fig.~\ref{figdel} displays the $DAF$ (shifted by + 417 
d; dashed line) and the $DCF$ (solid line) for $\alpha$ = 9 d. Surprisingly, the $DAF$ 
should be shifted to the right by a few days to optimally match the $DCF$. To assess the 
significance of this extra delay (excess lag with respect to 417 d), we analysed in detail the delay 
distributions based on Monte Carlo simulations (see above). We find that the LRT $r$-band main 
fluctuations with $S/N$ = 9.5 lead to $\Delta t_{BA}$ = 420.6 $\pm$ 1.9 d (1$\sigma$ interval).
Moreover, $\Delta t_{BA}$ is longer than 418.5 d at 91--98\% confidence levels, depending on the
value of $\alpha$. We can therefore state that chromaticity in $\Delta t_{BA}$ is detected at about 
the 2$\sigma$ level. 

\begin{figure}
   \centering
   \includegraphics[angle=-90,width=7cm]{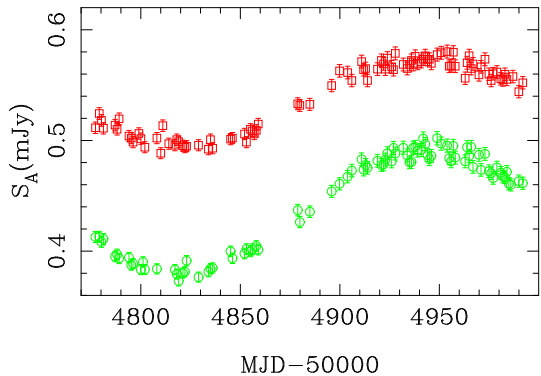}
   \caption{Two-colour light curves of Q0957+561A. These curves correspond to slightly extended 
   versions of the LRT main fluctuations of the A image: $g$ band (circles) and $r$ band 
   (squares).}
   \label{figagr}
   \end{figure}

This chromaticity in $\Delta t_{BA}$ is supported by interband delays for the two quasar images. 
From the LRT main fluctuations in the $g$-band and $r$-band fluxes of \object{Q0957+561B}, we 
infer $\Delta t_{rg} >$ 2.5 d at the 97\% confidence level ($\alpha$ = 9 d). A very accurate, 
1$\sigma$ delay $\Delta t_{rg}$ = 4 $\pm$ 1 d was also obtained from these data \citep{Gil12}. 
However, $\Delta t_{rg} <$ 2.5 d at the 98\% confidence level ($\alpha$ = 9 d) for 
\object{Q0957+561A} data. This last constraint is derived from densely sampled fluctuations 
in both optical bands (see Fig.~\ref{figagr}), which are slightly extended versions of the LRT 
main variations in the top panel of Fig.~\ref{figopt}. It seems that previous claims of a four-day
interband delay for the two images \citep{Col01,Sha08} were not accurate. Difficulties with 
relatively low $S/N$ values, monitoring gaps and other factors prevented \citet{Col01} from 
separating a four-day delay for B from an one-day delay for A, and did not allow us to accurately 
determine $\Delta t_{rg}$ for A. Although we adopted an 1$\sigma$ interval of 4 $\pm$ 2 d, some 
1$\sigma$ lower limits in Table 2 of Paper I are equal or close to 1 d. 

   \begin{figure*}
   \centering
   \includegraphics[angle=-90,width=7cm]{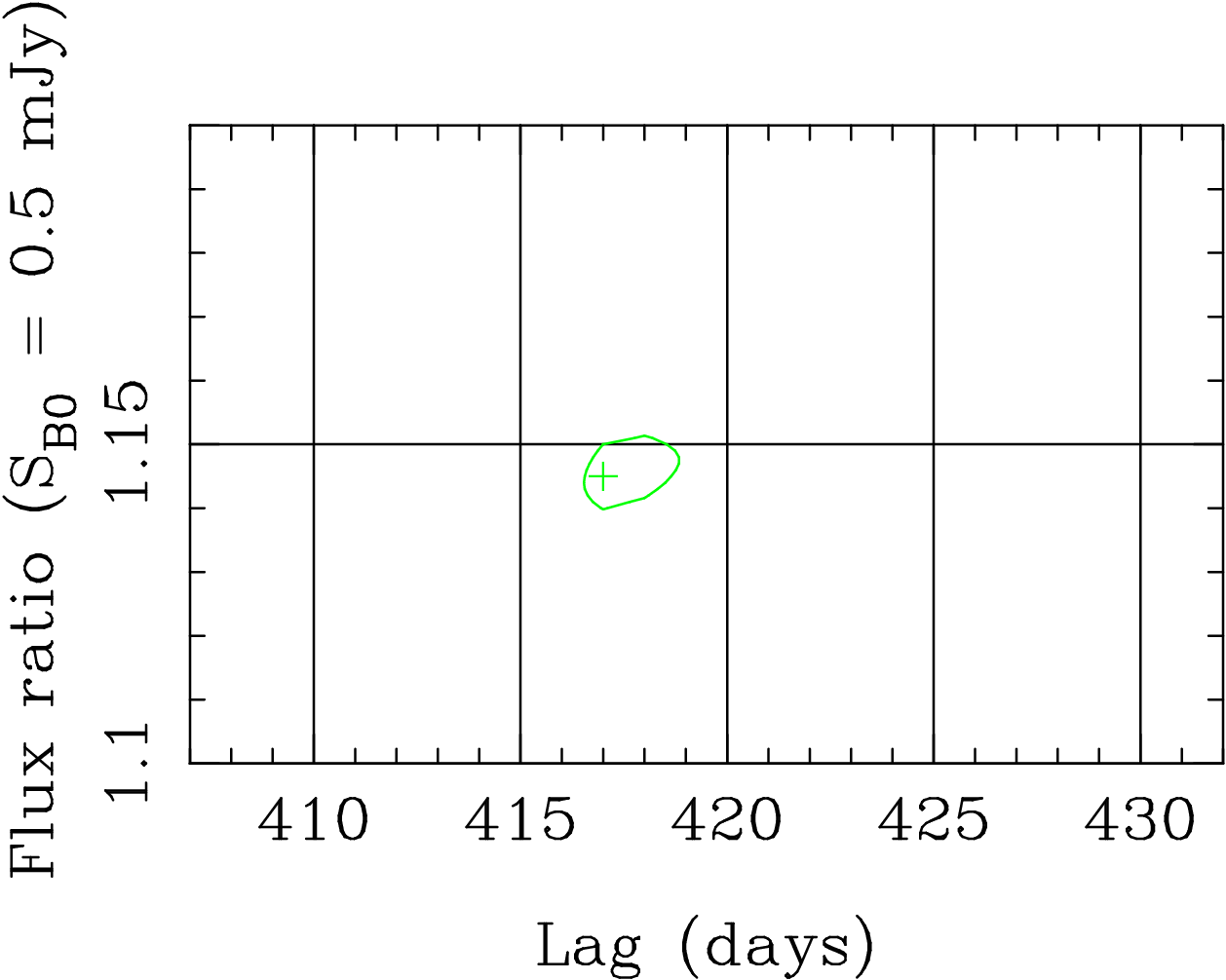}
   \hspace{1cm}
   \includegraphics[angle=-90,width=7cm]{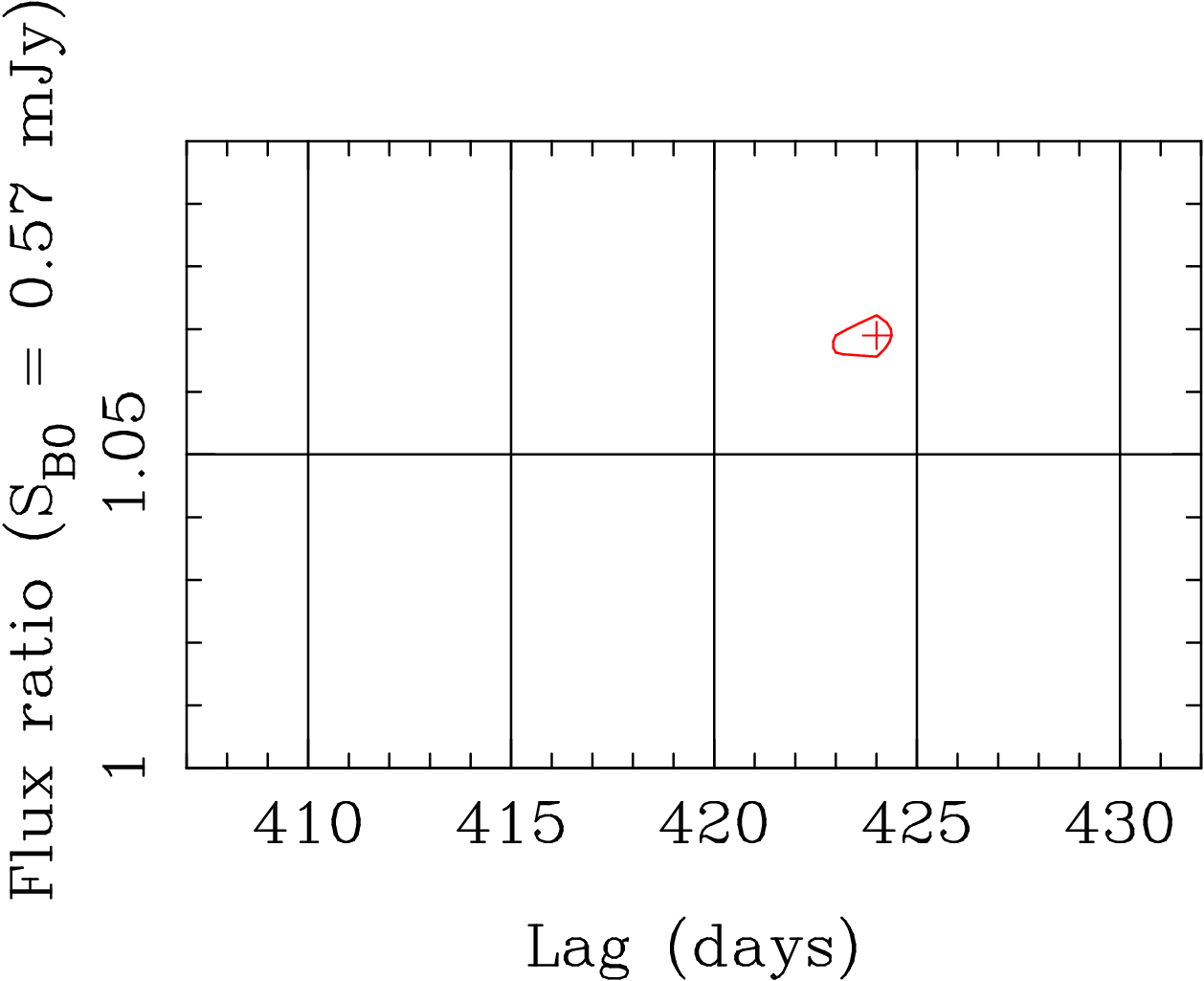}
   \includegraphics[angle=-90,width=7cm]{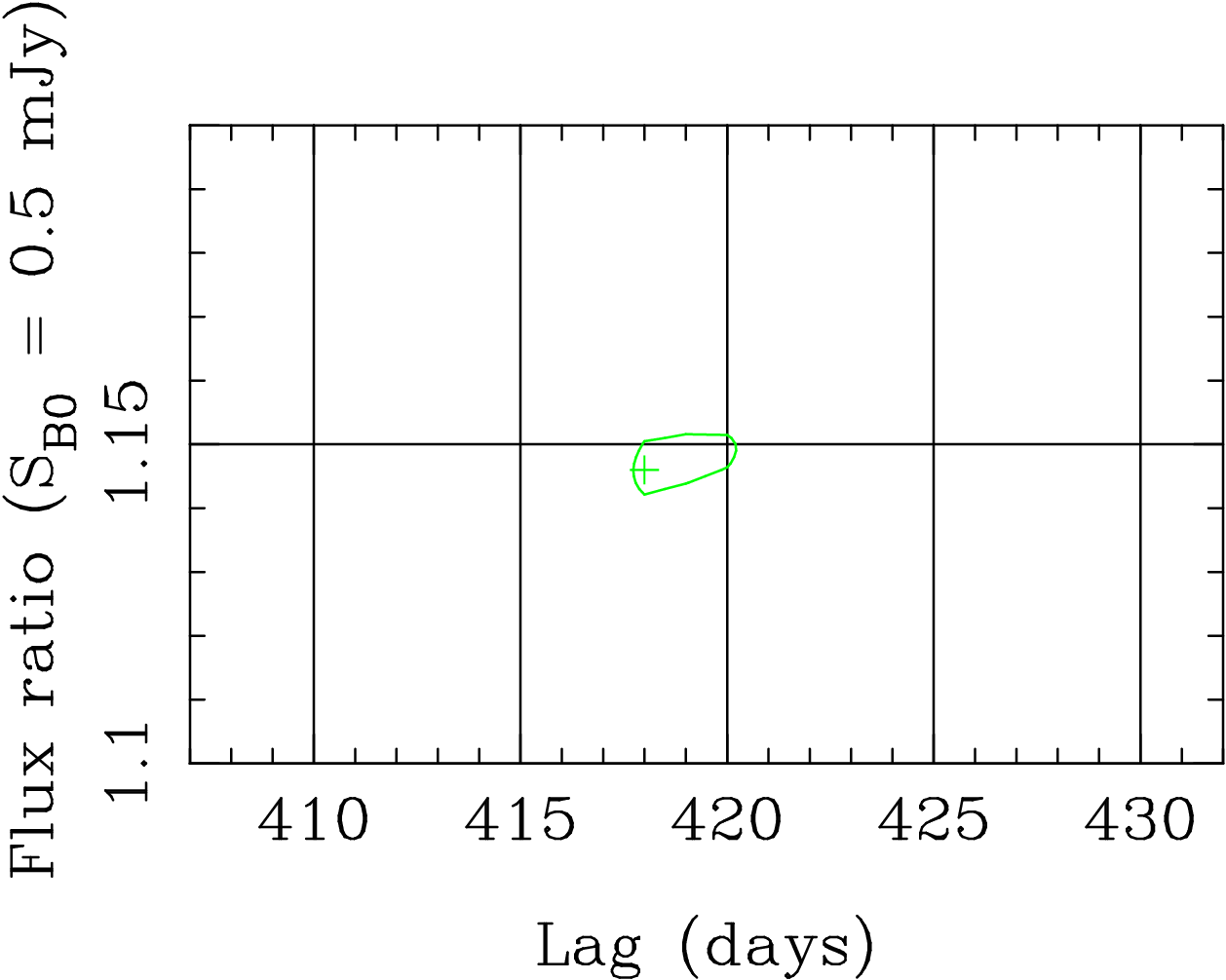}
   \hspace{1cm}
   \includegraphics[angle=-90,width=7cm]{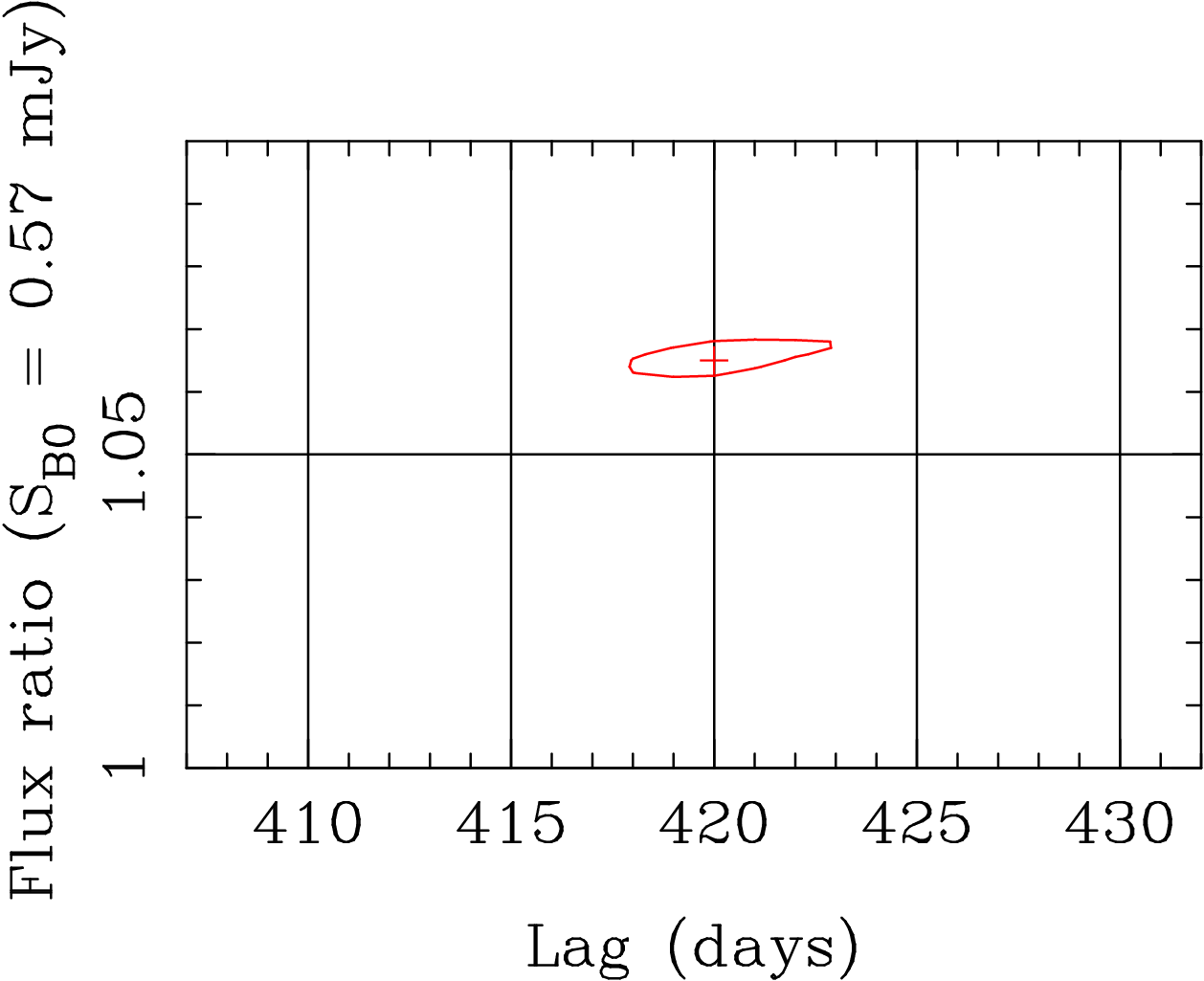}
   \caption{Best solutions and 2$\sigma$ contour lines from the $\chi^2$ technique. {\it Top 
   left panel}: $\alpha$ = 3 d ($g$ band). {\it Top right panel}: $\alpha$ = 3 d ($r$ band). 
   {\it Bottom left panel}: $\alpha$ = 9 d ($g$ band). {\it Bottom right panel}: $\alpha$ = 9 d 
   ($r$ band).}
   \label{figchi}
   \end{figure*}

The AB cross-correlation function is very sensitive to standard microlensing variability, 
i.e., uncorrelated variations in the two quasar images. If the LRT main fluctuations (left 
panels of Fig.~\ref{figdel}) would be affected by microlensing, then their autocorrelation and 
cross-correlation functions would have different shapes, and the cross-correlation peaks would 
not reach a maximum value of 1 \citep{Goi98}. However, these microlensing imprints are not seen 
in the right panels of Fig.~\ref{figdel}. Although slow microlensing was detected in light 
curves of several lensed quasars \citep[e.g.,][]{Gay05,Foh07,Sha09,Eul11}, typical gradients are
too small to play a role in brightness records over relatively short time segments. For example, 
\citet{Hai12} used LQLM I data and more recent measurements from the United States Naval 
Observatory (USNO) to study the $r$-band flux ratio of \object{Q0957+561}. In this analysis 
conducted in parallel to ours, the authors report on the possible existence of a microlensing 
gradient of 0.016 mag yr$^{-1}$ in the $r$ band (see, however, Section 4). Such a low gradient 
would produce an extrinsic variation of only 8 mmag over a six-month period, which is very much 
lower that the intrinsic signal that we find, and even lower than the noise level in our 
$r$-band data. Hence, the discussion throughout this paragraph indicates that standard 
microlensing is basically absent from the selected light curves and accordingly does not perturb 
the time delay estimates.

Despite the robustness of non-parametric methods based on correlation functions, we also 
considered a $\chi^2$ (parametric) technique to determine the time delays between the two images 
\citep[e.g.,][]{Kun97,Ull06}. This $\chi^2$ minimisation\footnote{We indeed minimised 
$\chi^2$/dof, with 'dof' being the degrees of freedom} allows us to check the quality of the 
parametric model for relating the shapes of $S_A$ and $S_B$. The simplest model consists of a 
constant flux ratio, i.e., $S_B(t) = (B/A) S_A(t-\Delta t_{BA})$, where $B/A$ is a constant. 
However, there is evidence for a $B/A-S_B$ correlation (see details and a discussion of this 
flux ratio behaviour in Sect. 4--6), therefore we assumed an observationally motivated model: $S_B(t) = 
(B/A) S_A(t-\Delta t_{BA})$, $B/A = (B/A)_0 [S_B(t)/S_{B0}]^{\beta}$, where $(B/A)_0$ is the 
flux ratio at the reference flux $S_{B0}$ and $\beta$ is the power-law index. Apart from the
time delay $\Delta t_{BA}$, this scheme involves two additional free parameters $(B/A)_0$ and 
$\beta$, which are used to link shapes. We do not know what the true way is to link $S_B$ to the 
correlated time evolution of $S_A$. A power-law flux ratio is only one option among a variety of 
possible models, and therefore, our $\chi^2$ results should be taken with caution. To compare $S_A$ 
and $S_B$, we also used bins in A with semisize $\alpha$. 

In the top panels of Fig.~\ref{figchi}, we display $\Delta t_{BA}$-$(B/A)_0$ maps including 
our best solutions for $\alpha$ = 3 d (crosses) and their associated 2$\sigma$ contour lines. 
The $g$-band and $r$-band results are shown in the left and right panels, respectively. We 
obtained reduced chi-square values close to 1 ($\chi^2$/dof $\sim$ 0.9), and two disjoint delay 
intervals around 417 d ($g$ band) and 424 d ($r$ band). These $g$-band and $r$-band delay 
intervals are separated by 4 d, whereas the difference between the best solutions is 7 d. Thus 
we find that the chromaticity in $\Delta t_{BA}$ is more pronounced than that from the 
$\delta^2$ method (see above). Although the results for $\alpha$ = 3 d support a significant 
chromaticity of the delay, other values of $\alpha$ produce overlapping delay intervals. For 
example, if we take $\alpha$ = 9 d, the best solutions are characterised by $\chi^2$/dof $\sim$ 
1.2--1.5. These appear in the bottom panels of Fig.~\ref{figchi} (crosses). We also show the 
2$\sigma$ contour lines around 418 d ($g$ band; bottom left panel) and 420 d ($r$ band; bottom 
right panel). For $\alpha$ = 9 d, both delay intervals overlap with each other, and the 
parametric technique does not separate the two delays in the two optical bands. However, the 
production of "excessive chromaticity" or achromaticity is not surprising, since the degeneracy 
between the two shape parameters and the delay likely prevents accurate/reliable $\chi^2$-based 
delay measurements.   

Hereafter, we consider the self-consistent solution for the delays: (a) $\Delta t_{BA}$ = 417 d 
in the $g$ band, (b) $\Delta t_{BA}$ = 420 d in the $r$ band, (c) $\Delta t_{rg}$ = 4 d for the 
B image, and (d) $\Delta t_{rg}$ = 1 d for the A image. Our solution agrees with the discussion 
in the previous paragraphs of this section. We also note that several studies of $\Delta t_{BA}$ 
in the red part of the optical spectrum favoured delays above 417 d that are only marginally 
consistent with the APO 2$\sigma$ interval in the $g$ band \citep[e.g.,][]{Ser99,Ova03a}. This 
discrepancy was not originally associated with a chromatic delay between images, but with less 
quasar variability and greater contamination from the lensing galaxy in red filters, the 
existence of multiple achromatic delays in long-term light curves, etc. 

\section{Two-colour flux ratio over the 2000--2010 decade}

The spectral behaviour and the long-term evolution of the delay-corrected flux ratio $B/A$ has 
attracted increasing attention in the first decade of this century 
\citep[e.g.,][]{Ref00,Osc02,Ova03b,Goi05a,Goi05b}. The 1999--2000 Hubble Space Telescope (HST) 
spectra of \object{Q0957+561} indicated the chromaticity of $B/A$ at optical continuum 
wavelengths \citep{Goi05a}. At the average wavelengths of the $g$ and $r$ bands, the HST data in 
Fig. 1 of \citet{Goi05a} lead to 1$\sigma$ intervals $B/A$ = 1.10 $\pm$ 0.01 ($g$-band) and 
$B/A$ = 1.04 $\pm$ 0.02 ($r$ band). It can be also demonstrated that the \hbox{C\,{\sc iii]}} 
($\lambda$1909) and \hbox{Mg\,{\sc ii}} ($\lambda$2798) emission lines only slightly influence 
the estimation of the optical continuum flux ratio from the $g$ and $r$ broad filters, 
introducing a small bias of $(B/A)_{\rm cont}/(B/A)_{\rm cont+line} \sim$ 1.01. Apart from 
spectral analyses, time domain studies suggested the constancy of the flux ratio between 1987 
and 2000 \citep[$R$ band; e.g.,][]{Osc02}, and then from 2000 to 2007 ($g$ and $r$ bands; Paper 
I). Here, the new LRT data allow us to discuss the $gr$ flux ratio in the 2007--2010 period, as 
well as to compare it with the HST two-colour ratio in 2000.

To evaluate the flux ratio in the $g$ band, one should compare the light curve of the B image 
and the fluxes of the A image shifted by + 417 d. For example, the \object{Q0957+561B} fluxes 
between day 3649 and day 3894 have a very short counterpart in the original light curve of 
\object{Q0957+561A} before day 3477 (see the circles in both panels of Fig.~\ref{figopt}). The 
counterpart only consists of three data points, and we did not calculate the $g$-band flux ratio 
over days 3649--3894. We also emphasize the absence of a counterpart in the $r$-band fluxes of A
for this time segment of B. The first useful time segment of B covers days 4010--4249, and it is 
labelled TS1. The fluxes of B in TS1 have a relatively long overlap with fluxes of A shifted 
by + 417 d (see the middle and bottom left panels of Fig. 7 in Paper I). We removed two data 
points from the overlapping record of A because these fluxes are affected by atmospheric and/or
instrumental problems (see Paper I for details). The other useful time segments of B are: TS2 
(from day 4498 to day 4627), TS3 (from day 4777 to day 4992) and TS4 (from day 5107 to day 
5373). 

\setcounter{table}{2}
\begin{table}
\begin{minipage}[t]{\columnwidth}
\caption{Flux ratio in the $g$ band.}
\label{tabfrg}
\centering
\renewcommand{\footnoterule}{}  
\begin{tabular}{lcccc}
\hline\hline
Time segment\footnote{See main text.} & $\alpha$\footnote{Bin semisize.} (d) & Best fit & 
$\chi^2$/dof\footnote{dof = degrees of freedom.} & $B/A$\footnote{2$\sigma$ confidence interval
when $\chi^2$/dof $<$ 1.2.} \\                  
\hline
TS1	&	6	&	1.083	&	46.4/50	&	1.078--1.089\\
	&	9	&	1.083	&	53.0/54	&	1.078--1.088\\
TS2	&	6	&	1.141	&	40.7/37	&	1.135--1.146\\
	&	9	&	1.143	&	61.3/40	&	--		\\
TS3	&	6	&	1.089	&	41.0/44	&	1.085--1.094\\
	&	9	&	1.088	&	52.2/50	&	1.083--1.092\\
TS4	&	6	&	1.137	&	126.3/52	&	--		\\
	&	9	&	1.137	&	152.3/54	&	--		\\
\hline
\end{tabular}
\end{minipage}
\end{table}

   \begin{figure*}
   \centering
   \includegraphics[angle=-90,width=14cm]{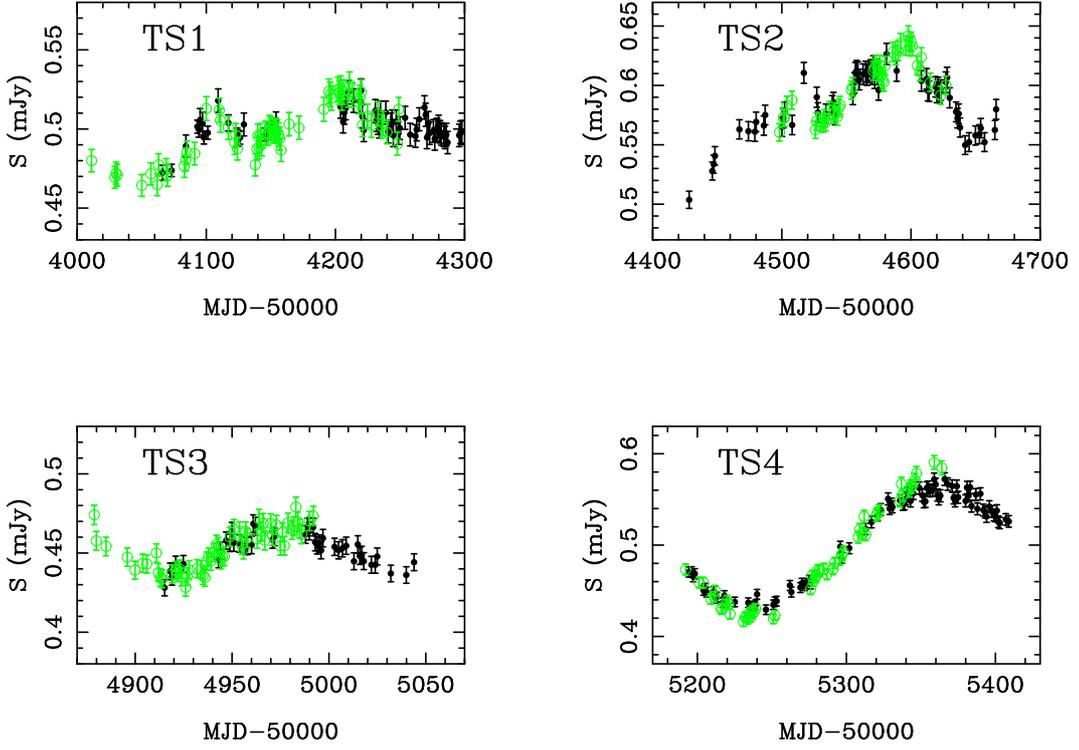}
   \caption{AB comparisons in the $g$ band. We show the overlapping periods between the light 
   curves of A (filled circles) and B (open circles), where the fluxes of A are shifted by + 
   417 d and are properly amplified (see main text). {\it Top left panel}: the overlap period for 
   the first time segment of B, i.e., TS1 (days 4010--4249). The A signal is amplified by 1.086. 
   {\it Top right panel}: the overlap period for TS2 (days 4498--4627). The A signal is 
   amplified by 1.14. {\it Bottom left panel}: the overlap period for TS3 (days 4777--4992). The 
   A signal is amplified by 1.086. {\it Bottom right panel}: the overlap period for TS4 (days 
   5107--5373). The A signal is amplified by 1.14.}
   \label{figfrg}
   \end{figure*}

   \begin{figure*}
   \centering
   \includegraphics[angle=-90,width=14cm]{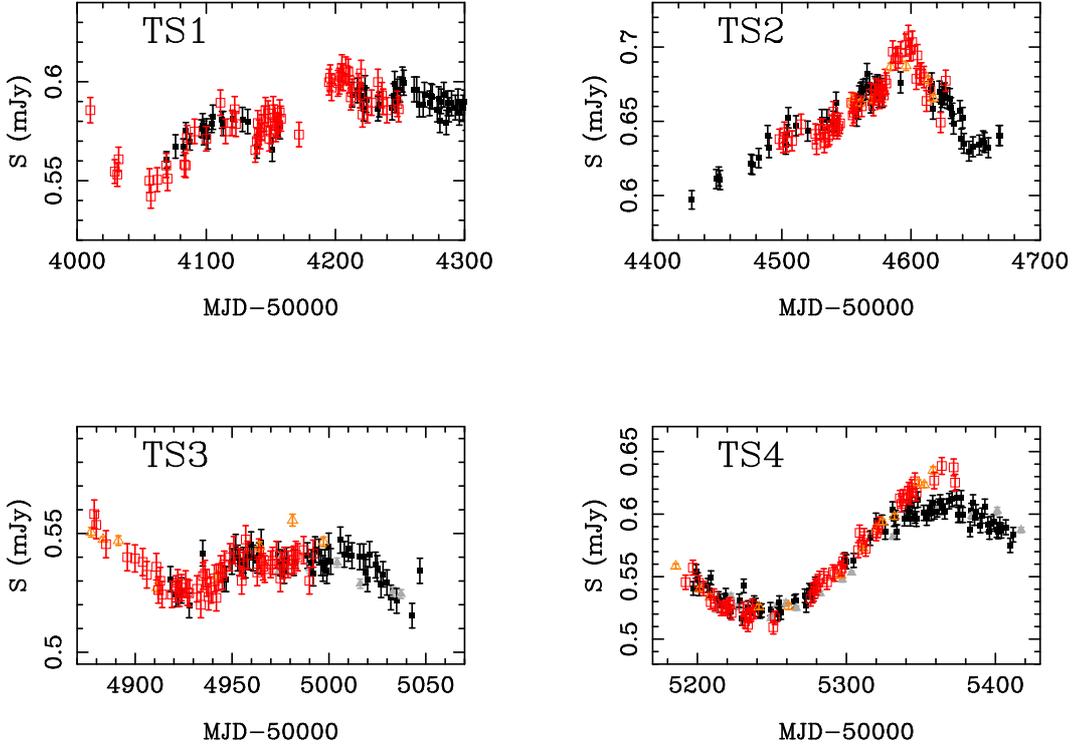}
   \caption{AB comparisons in the $r$ band. We used the original epochs and fluxes of B (open 
   squares), while the fluxes of A are shifted by + 420 d and properly amplified (filled 
   squares). Together with these LRT data, we also show the more poorly sampled USNO fluxes of A 
   (filled triangles) and B (open triangles). {\it Top and bottom left panels}: comparisons in 
   two periods of normal activity. The A signal is amplified by 1.019 (see main text). {\it Top 
   and bottom right panels}: comparisons in two episodes of violent activity. The A signal is 
   amplified by 1.057 (see main text).}
   \label{figfrr}
   \end{figure*}

\begin{table}
\begin{minipage}[t]{\columnwidth}
\caption{Flux ratio in the $r$ band.}
\label{tabfrr}
\centering
\renewcommand{\footnoterule}{}  
\begin{tabular}{lcccc}
\hline\hline
Time segment\footnote{See main text.} & $\alpha$\footnote{Bin semisize.} (d) & Best fit & 
$\chi^2$/dof\footnote{dof = degrees of freedom.} & $B/A$\footnote{2$\sigma$ confidence interval
when $\chi^2$/dof $<$ 1.2.} \\                  
\hline
TS1	&	6	&	1.023	&	50.5/46	&	1.019--1.027\\
	&	9	&	1.023	&	59.7/50	&	1.019--1.026\\
TS2	&	6	&	1.054	&	52.0/46	&	1.050--1.058\\
	&	9	&	1.056	&	80.5/51	&	--		\\
TS3	&	6	&	1.016	&	39.3/48	&	1.012--1.019\\
	&	9	&	1.016	&	34.1/50	&	1.012--1.019\\
TS4	&	6	&	1.060	&	160.5/54	&	--		\\
	&	9	&	1.060	&	169.0/57	&	--		\\
\hline
\end{tabular}
\end{minipage}
\end{table}

We used a $\chi^2$ minimization to find the flux ratio for each segment. In general, the shifted 
epochs of A do not coincide with the epochs of B, therefore we introduce bins in A around the epochs of 
B \citep[e.g.,][]{Ull06}. These bins have a semisize $\alpha$ of 6 or 9 d. In Table~\ref{tabfrg}
we give the best solutions and their reduced chi-square values, as well as some 2$\sigma$ 
intervals for $B/A$. A constant flux ratio $B/A$ = 1.086 works on both TS1 and TS3, and it agrees 
well with the corresponding HST ratio in 2000 (1.086 $\times$ 1.01 $\sim$ 1.10; see 
above). The AB comparisons for the time segments TS1 and TS3 are shown in the top and bottom 
left panels of Fig.~\ref{figfrg}, where we amplified the A signal by a factor of 1.086 (filled 
circles). Curiously enough, the simplest scenario (constant $B/A$) does work on TS2 and TS4, 
since most of the best solutions are associated with reduced chi-square values ranging from 1.5 
to 2.8. These best solutions $B/A \sim$ 1.14 also differ from the ratio for TS1, TS3 and the HST 
observing dates. Taking an amplification of 1.14 for the signal A in TS2 and TS4, both A and B 
signals are compared to each other in the top and bottom right panels of Fig.~\ref{figfrg} (A = 
filled circles and B = open circles). The flux ratio seems to reach "anomalous" values during 
episodes of violent activity (involving flux gradients $\geq$ 0.1 mJy/100 d; TS2 and TS4), while 
it remains lower and basically constant in other periods with flux gradients $\leq$ 0.05 mJy/100 
d. In the bottom right panel of Fig.~\ref{figfrg}, we also find evidence of a correlation 
between flux ratio and level of flux in TS4, i.e., $B/A <$ 1.14 at $S_B$ = 0.42 mJy, $B/A \sim$ 
1.14 at $S_B \sim$ 0.5 mJy and $B/A >$ 1.14 at $S_B$ = 0.59 mJy. This kind of 
correlation is not so evident in TS2.

In Table~\ref{tabfrr} we present our results in the $r$ band. For the two periods of normal 
activity in TS1 and TS3, a constant flux ratio $B/A$ = 1.019 can account for the flux gap 
between the light curve of B and the record of A shifted by + 420 d (see Table~\ref{tabfrr} and 
the left panels of Fig.~\ref{figfrr}). Although the adopted solution $B/A$ = 1.019 is only 
marginally consistent with the analyses in both periods, this should not cause suspicion of 
a possible decrease of $B/A$. The best-fit $\chi^2$/dof values for TS3 are 0.7--0.8, so the 
formal 2$\sigma$ intervals for this time segment only contain reduced chi-square values equal to 
or less than 0.9. Consequently, there are solutions $B/A \sim$ 1.021--1.022 with $\chi^2$/dof $\sim$ 1, 
which suggest that the flux ratio uncertainty for TS3 in Table~\ref{tabfrr} is underestimated. 
Taking the small perturbation by the \hbox{Mg\,{\sc ii}} ($\lambda$2798) line into account, we 
obtain a continuum flux ratio of $B/A \sim 1.019 \times 1.01 \sim 1.03$ at the average 
wavelength of the $r$ band (see above). This ratio agrees with the HST determination of $B/A$ at 
the same wavelength. For the two periods of violent activity (TS2 and TS4), a constant flux 
ratio does not convincingly explain the LRT observations. Moreover, the best solutions $B/A 
\sim$ 1.057 do not agree with those for TS1 and TS3. We display the AB comparisons for TS2 and 
TS4 in the right panels of Fig.~\ref{figfrr}, where the open squares trace the light curve of B,
and the delay-corrected and amplified fluxes of A are represented by filled squares. Once again, 
we find evidence of a $B/A-S_B$ correlation in the bottom right panel of Fig.~\ref{figfrr}, but 
this time in the $r$ band.

As we commented in Section 3, \citet{Hai12} found a slow gradient in the $r$-band flux ratio 
(in magnitudes) of \object{Q0957+561}. They used published LRT magnitudes together with new USNO 
data covering a general period similar to ours. However, we do not detect any long timescale 
drift in our analysis with only LRT data, and this discrepancy needs more attention. First, 
\citet{Hai12} adopted a LRT-USNO photometric offset of 14.455 mag, which seems to be biased in + 
0.025 mag when comparing LRT and USNO magnitudes at similar epochs. In their Fig. 2, the authors
derive the flux ratio in the second time-segment (days 4100--4200; it corresponds to our TS2) 
from a few differences $\Delta m_A$(LRT) - $\Delta m_B$(USNO). Because the USNO fluxes are likely 
underestimated in 0.025 mag, these differences should be enlarged until reaching the values for 
the fourth time-segment (days 4750--5000; TS4 in our framework). Second, the flux ratio in the
third time-segment (days 4500--4600; TS3 in our framework) is inferred from only a few USNO 
magnitude differences that include some outlier. In Fig.~\ref{figfrr} we compare the LRT and 
USNO fluxes using an unbiased LRT-USNO photometric offset of 14.43 mag, as well as turning HJD 
into MJD and magnitudes into mJy. The open triangles describe the USNO light curve of B, and the 
delay-corrected and amplified USNO fluxes of A are displayed as filled triangles. The time delay
and the amplifications are those obtained from the LRT data (see above). In general, the LRT and 
USNO data agree very well. However, there is a clear outlier in the 
USNO record of B for TS3. This noticeable deviation from the general trend means that $S_B$ is 
overestimated, and therefore, its associated magnitude should be increased by a certain amount. In 
Hainline et al.'s scheme, this would lead to a lower $\Delta m_A$ - $\Delta m_B$ value in 
the third time-segment, so the new cloud of magnitude differences would more closely resemble 
the cloud in the first time-segment (TS1 in our framework). Therefore, both the LRT and USNO 
data sets are consistent with an oscillating behaviour of $B/A$.

\section{Conclusions}

Our main conclusions are:
   \begin{enumerate}
      \item New LRT light curves of \object{Q0957+561A} and \object{Q0957+561B} in the $gr$ 
      bands during 2008--2010 show well-sampled, sharp intrinsic fluctuations with $S/N \sim$ 
      10. These extraordinary features allowed us to very accurately determine the $g$-band and 
      $r$-band time delays between both quasar images. The two time delays are inconsistent with 
      each other at the 2$\sigma$ level. More specifically, while we obtain a delay near to 417 
      d in the $g$ band, there is an extra delay of about three days in the $r$ band. This extra 
      delay cannot be attributed to low $S/N$ values, contamination from the main lensing 
      galaxy, or similar artifacts.
  	\item From the LRT two-colour photometry of \object{Q0957+561} during 2005--2010, we inferred
	the $g$-band and $r$-band delay-corrected flux ratio in four different time segments. The 
	flux ratio $B/A$ has an oscillating behaviour in both optical bands, reaching higher 
	values in the two segments of violent activity and remaining lower in the other two 
	periods of normal activity. These normal activity periods are characterised by $g$-band 
	flux gradients $\leq$ 0.05 mJy/100 d, and $B/A$ in each band does not vary from period to 
	period or within a given period. The normal $g$-band and $r$-band ratios are also 
	consistent with the HST ratios in 2000 at the average wavelengths of the $g$ and $r$ 
	bands. For the two episodes of violent activity (showing $g$-band flux gradients $\geq$ 
      0.1 mJy/100 d), the flux ratio in each band is similar in both segments, but it seems to 
      be correlated with the intra-segment flux level. 
   \end{enumerate}

\section{Discussion and future work}

The optical continuum of \object{Q0957+561A} is plausibly affected by a dense dusty cloud inside 
the cD lensing galaxy at $z$ = 0.36 \citep{Goi05a,Goi05b}. Because the light propagation time in 
the intervening medium is expected to increase with decreasing wavelength \citep[chromatic 
dispersion; e.g.,][]{Bor99}, the presence of this substructure could be responsible for a three-day 
lag between $g$-band and $r$-band signals, and thus explain the observed chromaticity of the 
delay between images. A detailed discussion on the composition and size of the cloud along the 
line of sight to the A image is beyond the scope of this paper. If this scenario turns out to be 
true, it would be necessary to estimate the proper delay to obtain a refined delay-based 
determination of the Hubble constant $H_0$ \citep[e.g.,][]{Jac07,Fad10}. In addition, future 
multi-wavelength (optical) monitoring campaigns of other gravitationally lensed quasars may also 
lead to unexpected delays caused by substructures in non-local lensing galaxies, and thus, 
to improved estimates of lensing mass distributions and $H_0$.

There is at least one crude interpretation for the flux ratio anomaly during violent episodes in 
\object{Q0957+561}. The violent activity may be related to a strong outflow, inducing a 
significant polarization degree in the otherwise weakly polarised UV emission 
\citep[e.g.,][]{Beg87,Bel98}. For the A image, this polarised radiation would pass through a 
dust-rich region with alligned elongated dust grains, suffering from a higher extinction than 
that observed in periods of normal activity \citep[dichroism; e.g.,][]{Bor99}. The induced 
polarisation degree could increase with increasing activity of the central engine (flux level), 
so that more extinction would be observed for higher fluxes. Future polarimetric data in both 
normal and violent periods will be used to check this interpretation. 

A microlensing scenario is difficult to reconcile with observations of \object{Q0957+561} for 
the last 25 years. The analysis of the LRT light curves indicates the absence of uncorrelated 
variations in the two quasar images (standard microlensing), and only a slight $B/A$ increase 
occurs for the sharpest intrinsic events. To account for this flux ratio anomaly, one might 
invoke the possible existence of radial expansions of the accretion disc during violent episodes
\citep[a model of an accretion disc with a time-varying size has also recently been proposed 
by][]{Bla10}. The expanded sources would cover larger regions of the microlensing magnification 
pattern for the B image, and produce slight extra magnifications of that image. Although such an
exotic microlensing seems to work, the "excessive constancy" of $B/A$ over $\sim$ 25 years 
\citep[e.g.,][and this paper]{Osc02} calls this scenario into question. We think the next 
logical step should be to accurately study light curves covering more than 5--6 years. New 
1999--2005 IAC-80 data in the $R$ band\footnote{A. Oscoz provided us with the $R$-band frames 
taken with the IAC-80 Telescope in the 1999--2005 period, within the framework of the Instituto 
de Astrof\'isica de Canarias (IAC)-Universidad de Cantabria (UC) collaboration. These frames 
will be fully reduced in a near future.} together with the 2005--2010 LRT and 2008--2010 USNO 
data in the $r$ band will make up a 10-year variability database, whereas additional old 
$R$-band fluxes \citep{Ova03a,Ser99} and 2011--2012 frames in the $r$ band may contribute to a 
20-year baseline. 

\begin{acknowledgements}
We thank L. J. Hainline and C. W. Morgan for kind interactions regarding our respective 
photometric approaches and data interpretations. The authors also thank the anonymous referee 
for valuable comments that improved the manuscript. We acknowledge the staff of the Liverpool 
Robotic Telescope (LRT) for their dedicated support and development of the Phase 2 User 
Interface, which allows users to specify in detail the observations they wish the LRT to make. 
The LRT is operated on the island of La Palma by Liverpool John Moores University in the Spanish 
Observatorio del Roque de los Muchachos of the Instituto de Astrof\'{\i}sica de Canarias with 
support from the UK Science and Technology Facilities Council. This research has been supported 
by the Spanish Department of Science and Innovation grants AYA2007-67342-C03-02 and 
AYA2010-21741-C03-03 (GLENDAMA project), and University of Cantabria funds. 

\end{acknowledgements}

\end{document}